# MEMORYRANGER PREVENTS HIJACKING FILE_OBJECT STRUCTURES IN WINDOWS KERNEL

Igor Korkin, PhD
Security Researcher
Moscow, Russia
igor.korkin@gmail.com

**ABSTRACT**

Windows OS kernel memory is one of the main targets of cyber-attacks. By launching such attacks, hackers are succeeding in process privilege escalation and tampering with users' data by accessing kernel-mode memory. This paper considers a new example of such an attack, which results in access to the files opened in an exclusive mode. Windows built-in security features prevent such legal access, but attackers can circumvent them by patching dynamically allocated objects. The research shows that the Windows 10, version 1809 x64 is vulnerable to this attack. The paper provides an example of using MemoryRanger, a hypervisor-based solution to prevent such attack by running kernel-mode drivers in isolated kernel memory enclaves.

## 1. INTRODUCTION

Modern preemptive multitasking operating systems like Windows and UNIX-based systems have two modes of operation: user mode and kernel mode. These modes are supported by CPUs and make it possible to isolate code and memory data in these two modes.

Apart from OS kernel and drivers, kernel-mode memory includes a lot of sensitive data structures, which can be used by attackers. CPUs do not provide any security features to prevent illegal access to that memory. As a result, intruders can gain read- and write- access to the kernel-mode memory by installing malware drivers or by exploiting driver vulnerabilities.

To mitigate these threats Windows has issued several protection mechanisms: PatchGuard, Device Guard etc. but they protect only fixed memory regions. For example, PatchGuard detects the Microsoft drivers hijacking (Unknowncheats, 2019). These security features do not completely prevent access to the dynamically allocated data structures.

For example, by exploiting Microsoft CVE-2018-8120 (Rapid7, 2018-a) vulnerability an attacker "could run arbitrary code in kernel mode". By using recently published vulnerabilities CVE-2018-8611 (Rapid7, 2018-b) and CVE-2018-8170 (Rapid7, 2018-c) attackers can elevate process privileges even on the newest Windows 10.

Newest APT such as FruityArmor and SandCat applies zero-day vulnerabilities of Windows OS kernel components, such as Win32k.sys to elevate process privileges: CVE-2018-8453, CVE-2018-8589, CVE-2019-0797. During these attacks, intruders patch the fields of EPROCESS structure, which corresponds to the particular process.

This paper considers a new kernel-mode memory attack on FILE_OBJECT structures, which makes it possible to read and write the content of the files opened by drivers in an exclusive mode. As a result, attackers can illegally access opened local and network files, which were not permitted for sharing.

The remainder of the paper proceeds as follows.

Section 2 provides the details of this attack and shows that security features from Windows 10 do not prevent it.

Section 3 contains the details of adapting MemoryRanger to prevent this attack and demonstrates that this solution successfully prevents this attack.

Section 4 and Section 5 focus on the main conclusions and further research directions respectively.

## 2. HIJACKING FILE_OBJECT TO GET AN ACCESS TO THE FILE OPENED IN EXCLUSIVE MODE

This section describes the internals of filesystem routines in the kernel mode with and without sharing access. The details of how to gain a full access to the file opened in an exclusive mode will be given in the second part of this chapter.

### 2.1. Overview of File System Kernel-mode Routines

Windows drivers call the following routines during file operations:

- ZwCreateFile – to create (or open) a file;
- ZwReadFile/ZwWriteFile – to read and write the file content;
- ZwClose – to close the file handle and release system resources.

The detailed overview of all the parameters for the functions are in MSDN (2017). All the steps for creating a file and getting the file handle are presented by Tanenbaum (2015).

The first function ZwCreateFile takes the full file name, flags etc and returns a handle to a successfully opened file or otherwise it returns error status. During this operation, the I/O manager calls the Object Manager to look up the named file and to help it resolve any symbolic links to the file object (Easefilter, n.d.).

Object Manager calls Security Reference Monitor (SRM) to process security checks, see Figure 1.

According to Yosifovich, Ionescu, Russinovich, & Solomon (2017) SRM determines "whether a file's Access Control List (ACL) allows to access the file in the way its thread is requesting. If it does, the object manager grants the access and associates the granted access rights with the file handle that it returns". I/O Manager builds FILE_OBJECTS with the help from the Object Manager (Russinovich, 1997; Nagar, 1997-a). The handle is used in read and write operations as well as to close the file. The FILE_OBJECT structure is an internal OS structure, which plays the role of kernel equivalent of a handle. As a result, each opened file has two structures in memory: a handle and a FILE_OBJECT, see Figure 1.

The functions ZwReadFile\ ZwWriteFile take the handle obtained at the previous step to read and write the files content. According to the Fernandez E. B. and Sinibaldi J.C. (2003) during these operations, SRM is not involved, see Figure 1, and this vulnerability can be used by attackers.

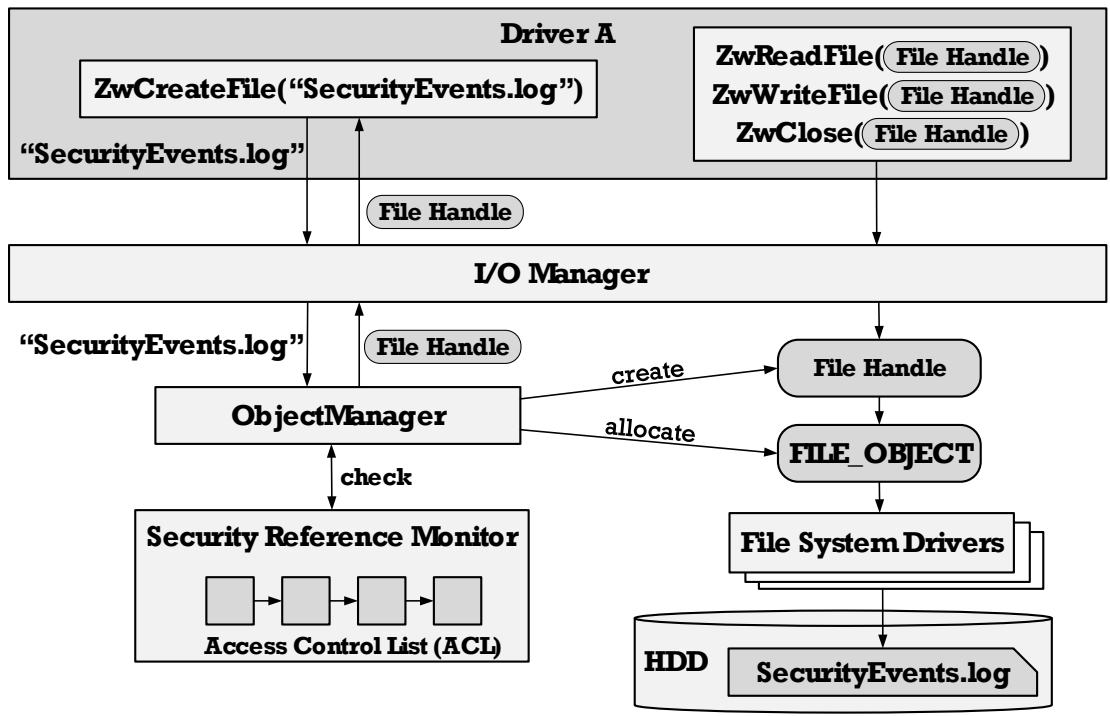

Figure 1 Internals of File System Routines in Windows Kernel

Finally, the ZwClose routine takes a handle to finish all writing operations and close an opened file. During this operation, SRM is not involved either.

### 2.2. Windows built-in security sharing

ZwCreateFile has a ShareAccess flag. This flag determines how the file is currently opened and also determines types of access allowed to proceed or deny with an error code of STATUS_SHARING_VIOLATION (Nagar, 1997-a).

Let us focus on the following scenario, see Figure 2. Driver A calls ZwCreateFile without sharing permission or with flag ShareAccess equals NULL. Object Manager successfully allocates FILE_OBJECT. Next the attacker's Driver calls the ZwCreateFile in order to gain a legal access to the file, which is already opened in an exclusive mode by DriverA. Object Manager returns the error status STATUS_SHARING_VIOLATION to the attacker's Driver, and prevents access to the file. As a result, the attacker fails to open such a file.

The present research reveals a vulnerability in File System Routines. The thing is that Object Manager addresses SRM only during ZwCreateFile call. Object Manager does not process any security checks during ZwReadFile and ZwWriteFile calls.

### 2.3. Analysis of FILE_OBJECT structure

Let us have a look at the details of FILE_OBJECT structure. As it was mentioned before, this structure is created by I/O Manager when a driver opens a file handle.

FILE_OBJECT structure includes about 30 fields, the detailed overview of all these files are presented by (McHoes, & Flynn, 2013) and (Nagar, 1997-b).

FILE_OBJECT fields partially duplicates the flags, which have been used during calling ZwCreateFile. For example, for the file opened by ZwCreateFile routine with flag ShareAccess, which equals NULL, the corresponding FILE_OBJECT structure has zero fields SharedRead and SharedWrite. The processed research shows that setting SharedRead and SharedWrite fields in the FILE_OBJECT does not allow the shared access to this file. The information about files sharing permission is also collected by SRM in the ACLs.

The following four FILE_OBJECT fields are used during read and write operations:

- Vpb;
- FsContext;
- FsContext2;
- SectionObjectPointer.

The Vpb field is initialized by the I/O Manager before sending a create or an open request to the file system driver. The Vpb field points to a mounted Volume Parameter Block (VPB), associated with the target device object.

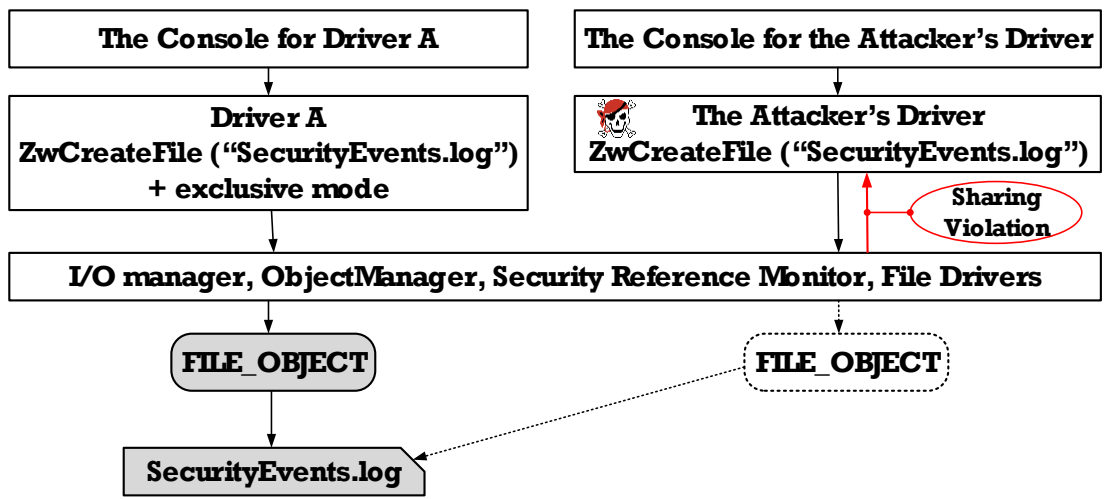

Figure 2. Windows OS prevents an illegal access attempt via calling ZwCreateFile to the file opened without sharing permissions

According to Nagar (1997-a) the FsContext, FsContext2, and SectionObjectPointer fields are initialized and maintained by the file system drivers and the NT Cache Manager.

FsContext points to the FSRTL_COMMON_FCB_HEADER structure, which has to be allocated by the file system or network driver.

FsContext2 field refers to the Context Control Block (CBB) associated with the file object.

SectionObjectPointer refers to a structure of type SECTION_OBJECT_POINTERS and stores file-mapping and caching-related information for a file stream.

These four fields are used in read and write files operations, which are processed without involving SRM and checking shared permission.

The key feature is that attackers can read these fields without any issues and use them to gain an access to the opened file. The details of this hijacked attack are below.

**2.4. Accessing the content of the file opened in an exclusive mode by hijacking its FILE_OBJECT**

Let us move on to the considered scenario, see Figure 3. In a similar way, DriverA has opened a target file in an exclusive mode. Object Manager successfully allocated the FILE_OBJECT structure to handle this file.

As it was mentioned before, the legal access to this file is blocked and the malware driver processes the following steps to gain the access illegally:

1. Calls ZwCreateFile routine to create a new file, e.g. with the name "hijacker.txt".
2. Calls ObReferenceObjectByHandle to get a pointer to the created FILE_OBJECT for the file hijacker.txt.
3. Finds FILE_OBJECT structure for the target file using the file name and walk through the Object Directory list (Probert, 2004; Pistelli, n.d.; Silberman, 2006; Microsoft. n.d.; GamingMasteR, 2009; Korkin & Nesterow, 2016; Fyyre, 2018; Abdalhalim, 2018).
4. Copies the following four fields from FILE_OBJECT for the target_file to the FILE_OBJECT for the file hijacker.txt:
    - Vpb;
    - FsContext;
    - FsContext2;
    - SectionObjectPointer.
5. Calls ZwReadFile/ZwWriteFile with the opened handle for the hijack_file to read and write the content of the target_file.

After processing these five steps, malware driver has achieved a full read and write access to the target file.

These manipulations were successfully tested on Windows 10, version 1809 x64. After waiting 10 hours, nothing happened, like appearing BSOD from PatchGuard, which is designed to prevent illegal memory modifications.

The experimental results show that a malware driver can gain full access to the opened file without sharing permission by hijacking its FILE_OBJECT and PatchGuard does not prevent this invasion.

The presented kernel attack is applicable to all modern Windows OSes since Windows NT 4.0. Windows components Object Manager and Security Reference Monitor involved in that attack, were first mentioned by Feldman (1993). And the first memory protection concept was developed in Multics system GE mainframe in 1965 by Corbató & Vyssotsky (1965).

To protect files content from being stolen and guarantee trusted computing the MemoryRanger hypervisor could be applied. The steps of adapting MemoryRanger to prevent this attack are in the next section.

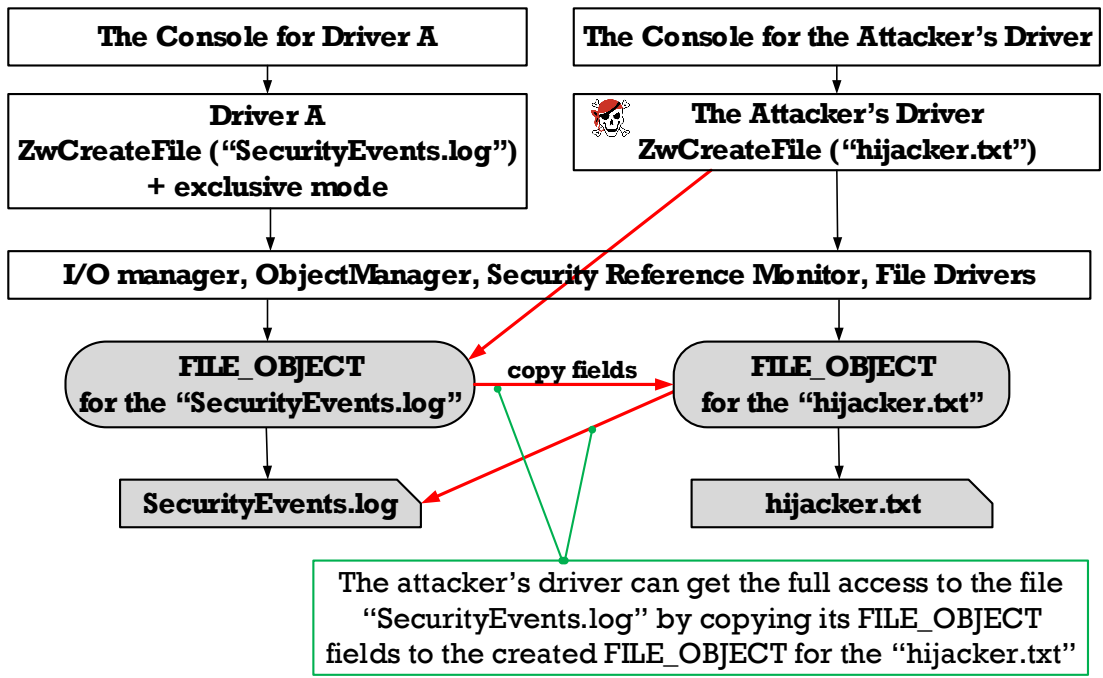

Figure 3. The attacker's driver hijacks the target FILE_OBJECT structure to gain an illegal access to the file opened without sharing permission

# 3. MEMORYRANGER PREVENTS FILE_OBJECT HIJACKING VIA MEMORY ISOLATION

MemoryRanger is an open-source solution designed to protect kernel-mode memory by creating isolated kernel enclaves and running drivers inside them (Korkin, 2018-a). MemoryRanger has flexible architecture, which makes it possible to extend it for the protection of new memory regions without any issues.

This chapter includes brief overview of main components of MemoryRanger and steps for adapting them to prevent FILE_OBJECT hijacking attack.

## 3.1. MemoryRanger architecture

MemoryRanger is a hypervisor-based solution and includes the following components, see Figure 4:

- A kernel-mode driver;
- DdiMon;
- MemoryMonRWX;
- Memory Access Policy (MAP).

The kernel-mode driver registers driver-supplied callback routines that are subsequently notified about various OS events, for example, about loading a new driver.

The next two components DdiMon and MemoryMonRWX leverage hypervisor facilities and use VT-x technology with Extended Page Tables (EPT) mechanism, provided by Intel CPU.

DdiMon is designed to monitor device driver interfaces and is able to hook kernel-mode API calls transparently for the OS.

MemoryMonRWX is able to track and trap all types of memory access: read, write, and execute.

Memory Access Policy (MAP) plays a role of an intermediary during memory access to the protected data and decides whether to block or allow access.

Initially MemoryRanger allocates the default EPT structure and puts all loaded drivers and kernel inside it. MemoryRanger traps loading of a new driver, then MemoryRanger allocates and configures a new EPT structure so that only this new driver and OS kernel are executed here. MemoryRanger isolates execution of drivers by switching between EPTs.

MemoryRanger hooks kernel API calls. The current version of MemoryRanger hooks ExAllocatePoolWithTag function to the protected newly allocated memory. Each time an isolated driver allocates memory, MemoryRanger updates all EPTs: the allocated memory buffer is accessible only for this driver, while all other EPTs exclude this memory. MemoryRanger skips the legal memory access attempts and prevents the illegal ones.

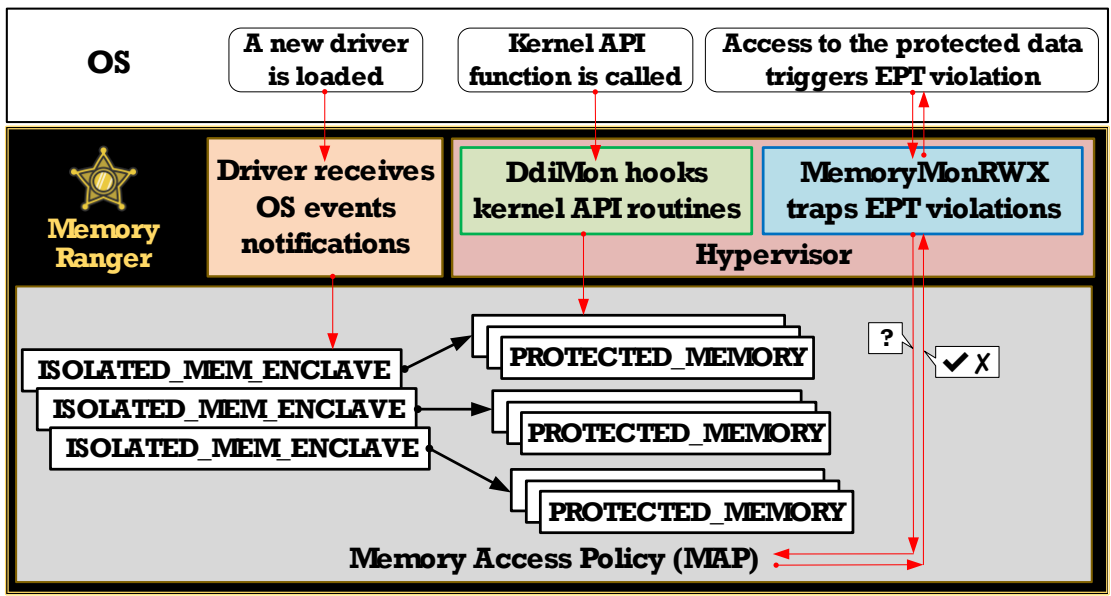

Figure 4. MemoryRanger architecture

### 3.2. Adapting MemoryRanger to protect FILE_OBJECT structures

To add a FILE_OBJECT support in MemoryRanger the modification of following components have been involved:

- DdiMon;
- MAP component.

The updated DdiMon hooks two file system-related routines ZwCreateFile() and ZwClose().

ZwCreateFile-callback routine processes the following:

1. Calls the original ZwCreateFile routine and checks whether the returned status is successful.
2. Checks whether the file has been created without sharing permission.
3. Checks whether the return address belongs to the protected drivers.
4. Gets the pointer to the allocated FILE_OBJECT by calling ObReferenceObjectByHandle.
5. Adds the FILE_OBJECT pointer and its size, which equals 0xB bytes to the protected memory region.

ZwClose()-callback routine processes the following:

1. Checks whether the return address belongs to the protected drivers.
2. Gets the pointer to the allocated FILE_OBJECT by calling ObReferenceObjectByHandle.
3. Deletes the FILE_OBJECT pointer and its size, which equals 0xB-bytes from the list of the protected memory region.

The MAP component algorithm processes access violation due to access to FILE_OBJECT structure in a similar way to the existing algorithm for processing access to the allocated memory pools.

The experimental results demonstrate that updated MemoryRanger has successfully protected FILE_OBJECT structures by preventing its hijacking without blocking legal access to FILE_OBJECT, see Figure 5. The source code of updated MemoryRanger and video demonstrations are here (Korkin, 2018-b).

The processing experiments on Windows 10 x64 have shown that developed updated MemoryRanger causes acceptable performance degradation.

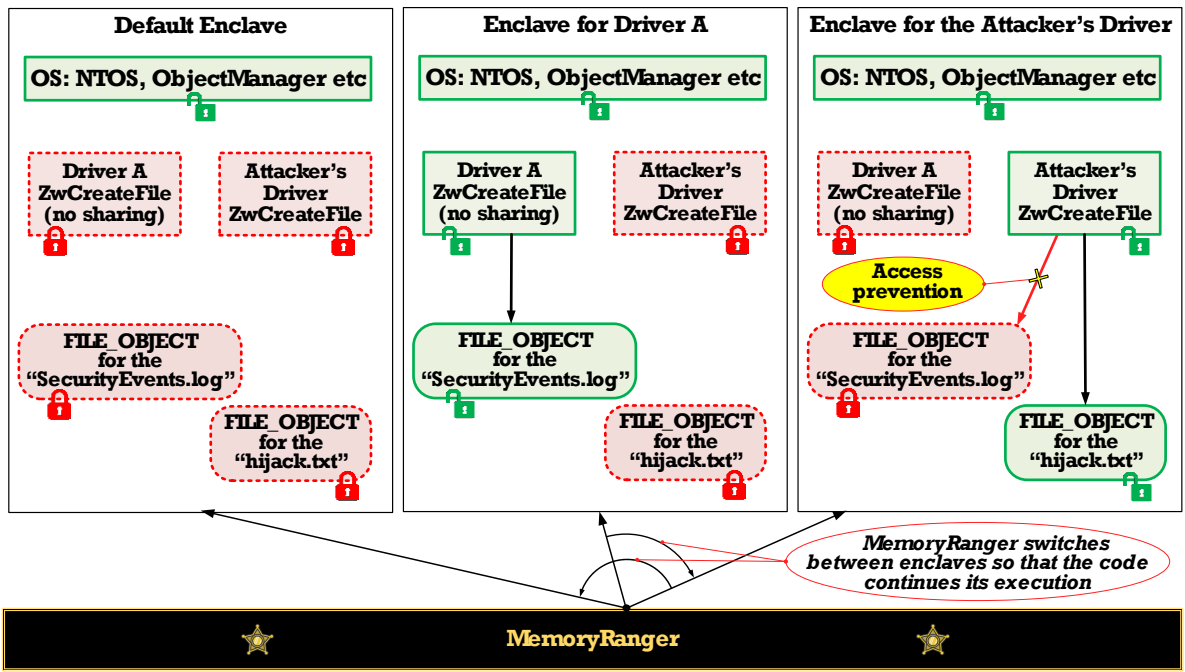

Figure 5. MemoryRanger prevents FILE_OBJECT hijacking by running drives into isolated kernel-mode memory enclaves

## 4. CONCLUSIONS

In summary:

1. Recently published kernel-mode exploits highlight the fact that dynamically allocated data structures in Windows OS kernel are becoming vulnerable.
2. The analyzed attack on FILE_OBJECT shows that the content of exclusively opened file can be tampered with.
3. The proposed update for MemoryRanger shows how to prevent hijacking attacks on FILE_OBJECT structures in kernel-mode memory.

## 5. FUTURE PLANS

### 5.1. Prevent Process Privilege Escalation

The analysis of recent kernel-mode vulnerabilities, such as CVE-2018-8120 (Rapid7, 2018-a), CVE-2018-8611 (Rapid7, 2018-b) and CVE-2018-8170 (Rapid7, 2018-c) shows that typically, vulnerable drivers do not access EPROCESS structures, but after exploitation, they tamper with process structures. For example during CVE-2018-8120 exploitation Win32k.sys driver directly accesses EPROCESS fields, but usually this driver does not communicate with this structure.

The current version of MemoryRanger traps the loading only of new drivers, moves only those to isolated enclosures. MemoryRanger skips already loaded drivers.

MemoryRanger is able to prevent this attack by deliberately running all loaded drivers in separate enclaves.

### 5.2. Protect ACL\DCL From Being Patched

Information about objects access rights is collected in the Access Control List (ACL). During ZwCreateFile routine call, the ObjectManager asks about required permission from Security Reference Monitor, which walks through the ACL to check the permissions. ACL includes access control entries (ACEs), which indicate what rights are granted to the object (Hewardt & Pravat, 2008; Swift, Brundrett, Dyke, Garg, Hopkins, Chan, Goertzel, & Jensenworth, 2002; Bosworth & Kabay, 2002; Datta, 2012; The NT Insider, 2006; The NT Insider 1999; Govindavajhala & Appel, 2006; MSDN, 2018; Russinovich, Ionescu & Solomon, 2012).

It seems promising to analyze the possibility of attacks on the ACL in order to deliberately change access mask and gain access to the target object.

MemoryRanger could be applied to provide the integrity of ACL and prevent these attacks in general.

## 6. REFERENCES

[1] Abdalhalim, A. (2018, January 14). A Light on Windows 10's "OBJECT_HEADER->TypeIndex". Retrieved from https://medium.com/@ashabdalhalim/a-light-on-windows-10s-object-header-typeindex-value-e8f907e7073a

[2] Bosworth, S., & Kabay, M. E. (2002). Operating System Security. Computer Security Handbook. 4th edition. John Wiley & Sons, Inc. New York, NY, USA

[3] Corbató, F. J., & Vyssotsky, V. A. (1965). Introduction and Overview of the Multics System. In Proceedings of the AFIPS Fall Joint Computer Conference, FJCC. Volume 27, Part 1. ACM, New York, NY, USA, pp. 185-196. DOI: http://dx.doi.org/10.1145/1463891.1463912

[4] Datta, A. (2012). Lecture 11: OS Protection and Security. CITS2230 Operating Systems. School of Computer Science & Software Engineering. The University of Western Australia. Crawley, Western Australia. Retrieved from http://teaching.csse.uwa.edu.au/units/CITS2230/handouts/Lecture11/lecture11.pdf

[5] Easefilter. (n.d.). Understand Windows File System File I/O. Retrieved from https://www.easefilter.com/Forums_Files/File_IO.htm

[6] Feldman, L (1993). Windows NT: The Next Generation. 1st edition. SAMS.

[7] Fernandez, E.B. & Sinibaldi, J.C. (2003). More patterns for operating systems access control. Paper presented at the Proceedings of the 8th European conference on Pattern Languages of Programs, EuroPLoP, pp. 381-398, DOI: https://doi.org/10.1.1.107.966

[8] Fyyre. (2018, November 10). WalkDirectory function. KernelDetective source code.

Retrieved from https://github.com/Fyyre/kerneldetective/blob/master/module.cpp

[9] GamingMasteR. (2009). Hidden Kernel Module (Driver) detection techniques. RCE forums. Retrieved from http://www.woodmann.com/forum/archive/index.php/t-12782.html

[10] Govindavajhala, S., & Appel, A. W. (2006, January 31).Windows Access Control Demystified. Princeton University. Retrieved from https://www.cs.princeton.edu/~appel/papers/winval.pdf

[11] Hewardt, M., & Pravat, D. (2008). Security. Advanced Windows Debugging. Addison-Wesley Professional.

[12] Korkin, I. (2018-a). Divide et Impera: MemoryRanger Runs Drivers in Isolated Kernel Spaces. In Proceedings of the BlackHat Europe Conference, London, UK. Retrieved from https://www.blackhat.com/eu-18/briefings/schedule/#divide-et-impera-memoryranger-runs-drivers-in-isolated-kernel-spaces-12668

[13] Korkin, I. (2018-b). MemoryRanger source code. GitHub repository. Retrieved from https://github.com/IgorKorkin/MemoryRanger

[14] Korkin, I., & Nesterow, I. (2016, May 24-26). Acceleration of Statistical Detection of Zero-day Malware in the Memory Dump Using CUDA-enabled GPU Hardware. Paper presented at the Proceedings of the 11th Annual Conference on Digital Forensics, Security and Law (CDFSL), Embry-Riddle Aeronautical University, Daytona Beach, Florida, USA, pp. 47-82 Retrieved from commons.erau.edu/adfsl/2016/tuesday/10

[15] McHoes, A., & Flynn, I. (2013). File Management. Windows Operating Systems. Understanding Operating Systems. Cengage India; 6th edition

[16] Microsoft. (n.d.). NtQueryDirectoryObject function. Microsoft Corporation. Retrieved from https://lacicloud.net/custom/open/leaks/Windows%20Leaked%20Source/wrk-v1.2/base/ntos/ob/obdir.c

[17] MSDN. (2017, June 17). Using Files in a Driver. Kernel-Mode Driver Architecture. Retrieved from https://docs.microsoft.com/en-us/windows-hardware/drivers/kernel/using-files-in-a-driver

[18] MSDN. (2018, May 5). ACCESS_ALLOWED_ACE structure. Retrieved from https://docs.microsoft.com/en-us/windows/desktop/api/winnt/ns-winnt-_access_allowed_ace

[19] Nagar, R. (1997, September-a). Windows NT File System Internals A Developer's Guide. O'Reilly Media. Retrieved from https://doc.lagout.org/operating%20system%20/Windows/Windows%20NT%20File%20System%20Internals%20-%20A%20Developer%27s%20Guide%20%281997%29.pdf

[20] Nagar, R. (1997, September-b). Fields in the File Object. Windows NT File System Internals A Developer's Guide. O'Reilly Media. Retrieved from https://doc.lagout.org/operating%20system%20/Windows/Windows%20NT%20File%20System%20Internals%20-%20A%20Developer%27s%20Guide%20%281997%29.pdf

[21] Pistelli, D. (n.d.). AntiMida 1.0. Retrieved from https://www.ntcore.com/files/antimida_1.0.htm

[22] Probert, D. (2004). Windows Kernel Internals Object Manager & LPC. Microsoft. Retrieved from http://i-web.i.u-tokyo.ac.jp/edu/training/ss/msprojects/data/04-ObjectManagerLPC.ppt

[23] Rapid7. (2018-a). Vulnerability & Exploit Database. Microsoft CVE-2018-8120: Win32k Elevation of Privilege Vulnerability. Retrieved from https://www.rapid7.com/db/vulnerabilities/msft-cve-2018-8120

[24] Rapid7. (2018-b). Vulnerability & Exploit Database. Microsoft CVE-2018-8611: Windows Kernel Elevation of Privilege Vulnerability. Retrieved from https://www.rapid7.com/db/vulnerabilities/msft-cve-2018-8611

[25] Rapid7. (2018-c). Vulnerability & Exploit Database. Microsoft CVE-2018-8170: Windows Image Elevation of Privilege Vulnerability. Retrieved from https://www.rapid7.com/db/vulnerabilities/msft-cve-2018-8170

[26] Russinovich, M. (1997). Inside NT's Object Manager. Compute Engines. Retrieved from https://www.itprotoday.com/compute-engines/inside-nts-object-manager

[27] Russinovich, M., Ionescu, A., & Solomon, D. (2012, March 15). Microsoft Windows Security. Microsoft Press Store. Retrieved from https://www.microsoftpressstore.com/articles/article.aspx?p=2228450&seqNum=3

[28] Silberman, P. (2006). FindObjectTypes function. Source code of FUTo_enhanced rootkit. Retrieved from http://read.pudn.com/downloads133/sourcecode/windows/system/568917/FUTo_enhanced/FUTo/Sys/Rootkit.c__.htm

[29] Swift, M.M., Brundrett, P., Dyke, C.V., Garg, P., Hopkins, A., Chan, S., Goertzel, M., & Jensenworth, G. (2002). Improving the Granularity of Access Control in Windows NT. Published in: Journal ACM Transactions on Information and System Security (TISSEC). Volume 5 Issue 4. pp. 398-437. DOI: https://doi.org/10.1145/581271.581273

[30] Tanenbaum, A. S. (2015). Modern Operating Systems. In 11.3.3 Implementation of the Object Manager. 4th Edition. Pearson Education.

[31] The NT Insider (1999). Keeping Secrets - Windows NT Security (Part I). The NT Insider. Vol 6, Issue 3. Retrieved from http://www.osronline.com/article.cfm?id=56

[32] The NT Insider (2006). In Denial - Debugging STATUS_ACCESS_DENIED. The NT Insider. Vol 13, Issue 2. Retrieved from http://www.osronline.com/article.cfm?article=459

[33] Unknowncheats (2019). Unknowncheats Driver object Hijacking. Retrieved from https://www.unknowncheats.me/forum/anti-cheat-bypass/317617-driver-object-hijacking.html

[34] Yosifovich P., Ionescu A., Russinovich M.E., & Solomon D.A. (2017). Chapter 7 Security. Windows Internals 7th edition. Microsoft Press. Redmond, Washington.